\documentclass[a4paper, amsfonts, amssymb, amsmath, reprint, showkeys, nofootinbib, twoside, floatfix, pre,superscriptaddress]{revtex4-2}

\usepackage[utf8]{inputenc}
\usepackage[left=23mm,right=13mm,top=35mm,columnsep=15pt]{geometry} 
\usepackage{amsmath}
\usepackage{amssymb}
\usepackage{esint}
\usepackage[unicode=true, pdfusetitle, bookmarks=true, bookmarksnumbered=false, bookmarksopen=false, breaklinks=false, pdfborder={0 0 1}, backref=false, colorlinks=false]{hyperref}
\usepackage{graphicx}
\graphicspath{ {./images/} }
\usepackage{xcolor}
\usepackage[shortlabels]{enumitem}

\makeatletter
\usepackage{makeidx}\usepackage{amsfonts}

\makeatother

\begin{document}

\title{Evidence of Replica Symmetry Breaking under the Nishimori conditions in epidemic inference on graphs}

\author{Alfredo Braunstein}
\affiliation{DISAT, Politecnico di Torino, Corso Duca Degli Abruzzi 24, 10129 Turin, Italy} 
\affiliation{INFN, Sezione di Torino, Turin, Italy}
\affiliation{Italian Institute for Genomic Medicine, IRCCS Candiolo, SP-142, 10060 Candiolo, TO, Italy.} 
\author{Louise Budzynski}
\affiliation{DI ENS, École Normale Supérieure, PSL, CNRS, INRIA}
\affiliation{Dipartimento di Fisica, Sapienza Università di Roma, Piazzale Aldo Moro 5, Rome 00185, Italy} 
\author{Matteo Mariani}
\affiliation{Department of Neurobiology, David Geffen School of Medicine, University of California, Los Angeles, Los Angeles, CA 90095,USA} 
\author{Federico Ricci-Tersenghi}
\affiliation{Dipartimento di Fisica, Sapienza Università di Roma, Piazzale Aldo Moro 5, Rome 00185, Italy} 
\affiliation{CNR-Nanotec, Rome unit and INFN, sezione di Roma1, Piazzale Aldo Moro 5, Rome 00185, Italy}
\newcommand{\com}{\textcolor{blue}}

\begin{abstract}
In Bayesian inference, computing the posterior distribution from the data is typically a non-trivial problem, which usually requires approximations such as mean-field approaches or numerical methods, like the Monte Carlo Markov Chain. Being a high-dimensional distribution over a set of correlated variables, the posterior distribution can undergo the notorious replica symmetry breaking transition. When it happens, several mean-field methods and virtually every Monte Carlo scheme can not provide a reasonable approximation to the posterior and its marginals. Replica symmetry is believed to be guaranteed whenever the data is generated with known prior and likelihood distributions, namely under the so-called Nishimori conditions. In this paper, we break this belief, by providing a counter-example showing that, under the Nishimori conditions, replica symmetry breaking arises. Introducing a simple, geometrical model that can be thought of as a patient zero retrieval problem in a highly infectious regime of the epidemic Susceptible-Infectious model, we show that under the Nishimori conditions, there is evidence of replica symmetry breaking. We achieve this result by computing the instability of the replica symmetric cavity method toward the one step replica symmetry broken phase. The origin of this phenomenon -- replica symmetry breaking under the Nishimori conditions -- is likely due to the correlated disorder appearing in the epidemic models.
\end{abstract}
\maketitle

\section{Introduction} 

Inference is the process of extracting a law, a relation, or a pattern by generalizing from some evidence or data. In Bayesian inference, this generalization is achieved by weighting a prior belief with the likelihood of data. In this way, a \textit{posterior} probability distribution on the hypothetical law is computed. Bayesian inference has been applied to a diverse set of scientific and technical fields, spanning from signal theory \cite{xu_target-oriented_2023,devasenapathy_transmission-efficient_2023}, artificial intelligence \cite{goldt_bayesian_2023,cui_bayes-optimal_2023}, computational biology and epidemiology \cite{altarelli_bayesian_2014,GhArBiZd23,BrBuMa23,braunstein_inference_2023,braunstein_small-coupling_2023}.

It is not surprising that several efforts are currently made to further understand Bayesian inference at a deep, fundamental level, with the introduction of simple models (as the planted Spin Glass \cite{ZdKr16}) devoted to discovering and understanding new properties of the posterior distribution \cite{RiSeZd19}.
Statistical physics heavily intersects with Bayesian inference, with a contact point being in the high dimensionality of the probability distributions studied in the two fields. For example, a phenomenon discovered in physics, that however applies to general high-dimensional probability distributions, is the replica symmetry breaking (RSB) transition. When this happens, the probability distribution becomes hard to sample from. Intuitively, a local Markov Chain Monte Carlo running for a time sub-exponential in the number of variables is unable to sample a representative set of instances from a RSB probability distribution. It is thus crucial to understand in which case the posterior can undergo such a transition.

It has been stated that a condition, named \textit{Nishimori condition}, would guarantee the absence of RSB in the posterior \cite{ZdKr16}. This conjecture has been proven to be exact in specific cases \cite{Montanari08,BaPa22}. The Nishimori conditions can be intuitively explained as the case in which the computation of the posterior is made when the prior and the likelihood are exactly known. Suppose, for example, the process of inferring a signal passing through a noisy channel; if the stochastic law of signal distortion due to noise and the distribution of the source are known, then the Nishimori conditions are met. 

When RSB is not present, then an algorithm named Belief Propagation (BP) should always converge in locally tree-like, infinite networks. However, a systematic non-convergence of BP was observed in the context of epidemic inference on the Susceptible-Infectious (SI) model by \cite{GhArBiZd23} under Nishimori conditions. Initially disguised as a finite size effect, this phenomenon was, however, observed in \cite{BrBuMa23} too, where a replica symmetric (RS) cavity method approach allowed to study the infinite size limit of the same SI model under the Nishimori conditions. RSB was thus conjectured for the first time as a possible explanation of the phenomenon, but little support was given to the claim. Here we consider again a sub-case of the model in \cite{GhArBiZd23} and \cite{BrBuMa23} and we show that RSB arises in the posterior distribution under Nishimori conditions.

\begin{figure}
	\centering
	\includegraphics[width=\columnwidth]{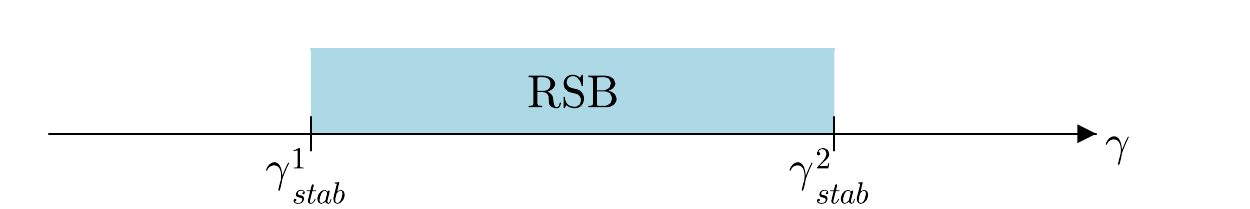}
	\caption{The RSB phase exists for $\gamma\in[\gamma_{\rm stab}^1,\gamma_{\rm stab}^2]$, where $\gamma$ is the probability of being infected at the initial time (probability of being a seed). In Sec.~\ref{subsec:instability_RS} we obtain the values of the two instability threshold --- $\gamma_{\rm stab}^1=0.00542$, $\gamma_{\rm stab}^2=0.0152$ --- for observation time $T_{obs}=8$, on random regular graphs of degree $3$, and transmission rate $\lambda=1$ (deterministic spreading dynamics).}
	\label{fig:phase_diagram}
\end{figure}

In this work, we study the RS cavity equations again, unveiling a local instability of the RS solution towards the space of RSB solutions. Specifically, this is called Kesten-Stigum instability in the context of tree reconstruction \cite{KeSt66,MoPe03}, or the deAlmeida-Thouless transition for mean-field spin glasses \cite{AlTh78}.
We precisely locate the stability threshold and thus the limits of this RSB phase in parameters space, in the thermodynamic (infinite size) limit, see Fig.~\ref{fig:phase_diagram}.
Additionally, we also provide an instability study of the BP solution found in finite-size instances, and locate the limits of its unstable phase. 

In contrast to other classical statistical inference problems such as the planted spin glass \cite{ZdKr16}, the epidemic inference problem presents the particularity of having {\it correlated} disordered, where the disordered variables are the observations given in the inference process.
In two recent papers \cite{Nishimori24, Nishimori25}, the role of correlations in the disorder is investigated in a specific setting of spin glass models, where it is shown that the traditional argument against an RSB phase on the Nishimori line does not apply.
These findings point towards the idea that correlated disorder might be at the origin of the RSB phase transition we unveil in epidemic inference.
More generally, it opens the way for further investigations on the role of correlated disorder in statistical inference problems, and for a better understanding of the mechanism responsible for symmetry breaking on the Nishimori line.

\section{Methods}
\subsection{Epidemic inference on networks}

\paragraph{SI model on graphs}
We consider the SI model of spreading on a graph $G=(V,E)$. It is a stochastic dynamical model in discrete time, in which the state of node $i\in V$ at time $t$ is represented by a variable $x_i^t\in\{S,I\}$.
At each time step, an infected ($I$) node $i$ can infect each of its susceptible ($S$) neighbors $\partial i=\{j\in V: (i,j)\in E\}$ with probabilities $\lambda_{ij}^t\in[0,1]$.
The dynamic of SI model is irreversible: a node $i$ can only undergo the transition $S\to I$. We can therefore describe an individual's trajectory by a single variable: its infection time $t_i$.
We assume that at time $t=0$, a subset of infectious nodes are initiated with an infection time $t_i=0$: $x_i^0=I$.
A realization of each individual's trajectories can be unequivocally expressed in terms of transmission delays $s_{ij}\in\{1,2,\dots,\infty\}$, sampled from a geometrical law: $w_{ij}(s_{ij}) = \lambda_{ij}(1-\lambda_{ij})^{s_{ij}-1}$.
Once the initial condition $\{x_i^0\}_{i\in V}$ and the set of delays $\{s_{ij},s_{ji}\}_{(i,j)\in E}$ are fixed, the set of infection times $\underline{t}=\{t_i\}_{i\in V}$ is the unique solution to the system of equations:
\begin{align}
\label{eq:equation_infected_times}
t_i=\delta_{x_i^0,S}\min_{j\in\partial i}\{t_j+s_{ji}\}
\end{align}

We consider homogeneous and time-independent trans\-mission rates $\lambda_{ij}^t=\lambda$ for all $t$ and $(i,j)\in E$. However, the analysis presented in this work can be straightforwardly extended to the inhomogeneous case.
We further simplify the model, considering a transmission rate $\lambda=1$, corresponding to a deterministic spreading process in which $s_{ij}=1$, such that at each time step each infected node infects its susceptible neighbors with probability $1$.
We also assume each node $i\in V$ to have the same probability $\gamma$, named {\it seed probability} of being infectious at time $t=0$.

The {\it prior probability} of infection times $\underline{t}=\{t_i\}_{i\in V}$ conditioned on the realization of delays $\{s_{ij},s_{ji}\}_{(i,j)\in E}$ and on the initial condition $\{x_i^0\}_{i\in V}$ can be written as
\begin{align}
P(\underline{t}|\{x_i^0\},\{s_{ij},s_{ji}\})&=\prod_{i\in V}	\psi^*(t_i, \underline{t}_{\partial i}, x_i^0, \{s_{ji}\}_{j\in\partial i}) 
\label{eq:prior:constraints_delay_times}
\end{align}
with $\underline{t}_{\partial i}=\{t_j, j\in\partial i\}$, and where $\psi^*$ enforces the constraint (\ref{eq:equation_infected_times}) on the infection times:
\begin{align}
\label{eq:constraint_infection_times}
\psi^*=\mathbb{I}[t_i=\delta_{x_i^0,S}\min_{j\in\partial i}\{t_j+s_{ji}\}]
\end{align}
with $\mathbb{I}[A]$ being the indicator function of the event $A$.

\paragraph{Inferring individuals trajectories from partial observations}

We assume that some information $\mathcal{O}=\{o_i\}_{i\in V}$ is given on the state of the nodes, at a given observation time $T_{\rm obs}$, by the result of $N$ independent clinical tests. 
The probability of observing $P(\mathcal{O}|\underline{t})$ factorizes on the set of tests:
\begin{align}
\label{eq:observations}
\begin{aligned}
P(\mathcal{O}|\underline{t})&=\prod_{i\in V}\rho(o_i|t_i) \quad \text{with}\\
\rho(o_i|t_i)&=\mathbb{I}[t_i\leq T_{\rm obs}]\delta_{o_i,1} + \mathbb{I}[t_i> T_{\rm obs}]\delta_{o_i,0}
\end{aligned}
\end{align}
Note that in \cite{BrBuMa23}, a more general setting was considered in which the observations could be noisy (introducing a positive false rate), and/or restricted to a subset of the nodes, or made at different time of the process. We restrict ourselves to {\it noiseless} observations of {\it all} individuals at the {\it same} observation time $T_{\rm obs}$, but the analysis presented in this paper could easily be generalized to other settings.

One aims to reconstruct unobserved information on the individual's trajectory.
In the Bayesian inference setting, the {\it posterior probability} $P(\underline{t}|\mathcal{O})$ of the infection times, given the observations, can be written in terms of the {\it prior probability} of individual trajectories and of the {\it likelihood probability} of observations: 
\begin{align}
\label{eq:posterior}
	P\left(\underline{t}\right|\mathcal{O})=\frac{P(\underline{t})P(\mathcal{O}|\underline{t})}{P(\mathcal{O})}
\end{align}

\paragraph{Nishimori conditions and Bayes optimal setting}

The Nishimori conditions (or Bayes optimal conditions) correspond to the setting where prior and likelihood parameters ($\gamma$ and $\lambda$ in our model) are {\it known} in the inference process.
When the Nishimori conditions are met, the Nishimori identities can be derived: for any function $f$ over two configurations of infection times, we have:
\begin{align}
\label{eq:Nishimori_id}
\mathbb{E}_{\mathcal{O}}[\langle f(\underline{t}_1,\underline{t}_2)\rangle_{\underline{t}_1,\underline{t}_2}] = \mathbb{E}_{\mathcal{O}}[\langle f(\underline{t},\underline{\tau})\rangle_{\underline{t},\underline{\tau}}]
\end{align}
In the l.h.s, the configurations $\underline{t}_1,\underline{t}_2$ are sampled independently from the posterior probability distribution (\ref{eq:posterior}), while in the r.h.s, $\underline{\tau}$ is the planted configuration (sampled from the prior), and $\underline{t}$ is sampled from (\ref{eq:posterior}). The average $\mathbb{E}_\mathcal{O}$ is across all the spreading processes resulting in the set of observations $\mathcal{O}$.
In particular, one can choose $f$ to be a general moment of the distribution of the overlap between the two configurations, thus concluding, by means of the Nishimori identity, that the overlap between a planted and an inferred configuration has the same distribution of the overlap between two inferred configurations. 

The typical argument for claiming that there is no RSB phase transition in the Bayes optimal setting is based on the assumption of self-averaging of the overlap between the planted and an inferred configuration. 
By the Nishimori identities, this would imply that the overlap between two inferred configurations is self-averaging, which can only be true in the Replica Symmetric phase.
However, in the case of epidemic inference, there is no guarantee that the overlap between planted and inferred configurations is self-averaging. 
In fact, our results in section \ref{subsec:instability_RS} unveil the presence of an RSB phase transition. 

Note that technically, the above argument does not exclude the presence of a {\it dynamic} \cite{BouchaudCugliandolo98} phase transition towards an 1RSB phase on the Nishimori line. A dynamic phase transition is indeed not captured by short-range correlation functions, such as (\ref{eq:Nishimori_id}), but only by the appearance of a specific kind of long-range correlations between variables, known as point-to-set correlations \cite{MoSe06}. These correlations forbid the rapid equilibration of stochastic processes satisfying detailed balance, hence the name dynamic.
The static properties of the posterior distribution (\ref{eq:posterior}) -- and in particular the overlap distribution -- are only affected by a {\it static} (or condensation) transition.
The stability analysis of the Replica Symmetric solution conducted in \ref{subsec:instability_RS} unveils an RSB phase, but does not allow us to identify the nature of the RSB phase, nor to exclude the existence of a dynamics 1RSB phase transition taking place in the RS phases. In order to clarify the latter point, one needs to solve the 1RSB equations. This is left for a future work.

\paragraph{Belief-Propagation Algorithm}

Several approaches exist to extract information from the posterior (\ref{eq:posterior}) and estimate the individuals trajectories on a given instance of this problem, including algorithms based on Belief Propagation (BP) \cite{AlBrDaLaZe14}.
BP is a method that allows to compute marginals of a probability measure whose underlying factor graph (graph of interaction between random variables) is a tree or can be approximated by a tree. In \cite{AlBrDaLaZe14}, it was shown how to apply this technique to sample the posterior probability distribution (\ref{eq:posterior}) in epidemic models. One introduces BP messages $\mu_{i\to j}$ on each directed edge of the contact graph $G$, solving the following BP equations:
\begin{align}
	\label{eq:BP_posterior}
	\mu_{i\to j} &= f^{\rm BP}(\{\mu_{k\to i}\}_{k\in\partial i \setminus j})
\end{align}
(see appendix \ref{app:BP_stab}, and equation (\ref{eq:BP_onegraph_app}) for a complete expression). 
On a given instance (i.e. a given contact graph $G$ and a given realization of the individual's trajectories and observations $\mathcal{O}$), these equations are solved numerically with an iterative procedure.
Once a solution is found, the marginal probabilities 
$$
P(t_i|\mathcal{O}) =\sum_{\{t_j\}_{j\neq i}} P(\{t_i\}_{i\in V}|\mathcal{O})
$$ 
can be estimated from the BP messages $\{\mu_{j\to i}\}_{j\in\partial i}$.
These marginals can in turn be used to compute optimal estimators for the reconstruction of the infected times $\{t_i\}_{i\in V}$, and therefore to solve the inference problem.
If the contact graph is a tree, BP is exact and the iterative procedure for solving (\ref{eq:BP_posterior}) converges to the correct and unique solution in a time growing at most logarithmically with the graph size.
Even when the contact graph is not a tree, BP can be used as an heuristic, and provides good estimates on locally tree-like graphs.
The convergence towards the correct solution is however not guaranteed, and depends on the correlation decay between distant variables due to the presence of long loops.
It was indeed observed in \cite{GhArBiZd23} that BP algorithm failed to converge towards a solution, for some values of the parameters defining the inference task.
In section \ref{subsec:instability_BP}, we provide a stability study of the BP equations (\ref{eq:BP_posterior}), and locate the region where BP is unstable. 

\subsection{Ensemble average}  

We define a single instance as a contact graph $G=(V,E)$, altogether with a realization of an epidemic spreading, which can be defined by specifying the {\it planted} infection times $\{\tau_i\}_{i\in V}$ of all the nodes. By averaging over an ensemble of instances, it is possible to compute, in the large size (thermodynamic) limit, the posterior probability (\ref{eq:posterior}), as a function of the epidemic parameters parameters such as the transmission rate $\lambda$, on the graph connectivity, or on the seed probability $\gamma$  \cite{BrBuMa23}. Following this approach, in this paper we focus on a specific sub-case, namely random regular graphs, with fixed degree $d$, and deterministic spreading dynamics ($\lambda=1$). In this regime we study the replica symmetry properties of the posterior distribution when changing the seed probability $\gamma$.

\begin{figure*}
	\centering
	\includegraphics[width=0.45\textwidth]{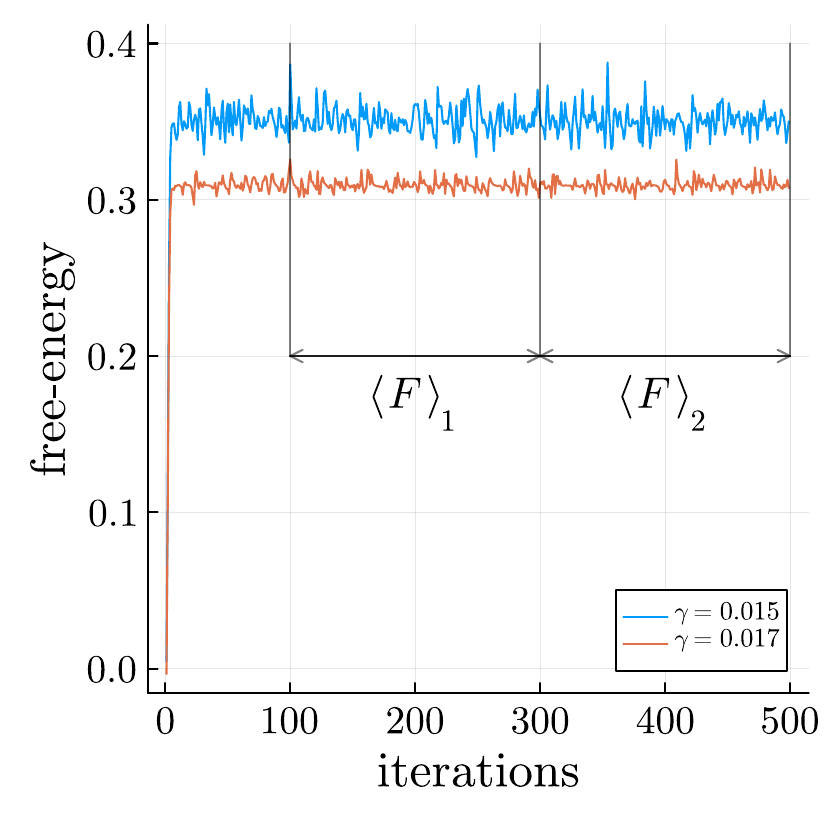} 	
	\includegraphics[width=0.45\textwidth]{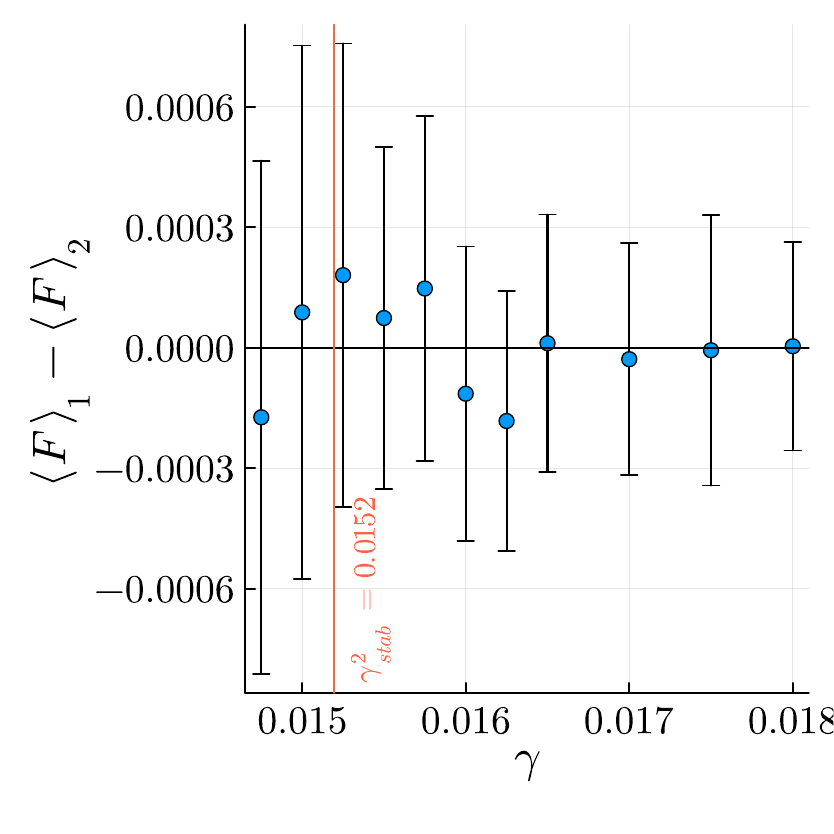} 
	\caption{Study of the RS fixed-point, for $\lambda=1$, $T_\text{obs}=8$, on random regular graphs with degree $d=3$. The population size is $\mathcal{N}=5\cdot10^4$. Left: Evolution of the RS free-energy (\ref{eq:RS_free_energy}) along iterations, for two values of $\gamma$: in the unstable phase $\gamma=0.015<\gamma_{\rm stab}^2=0.0152$ (upper blue curve), and in the stable phase $\gamma=0.017>\gamma_{\rm stab}^2$ (lower red curve). In both cases, a stationary regime is reached and the free-energy fluctuates around its mean value. Right: Convergence criterion for the RS fixed-point, as $\gamma$ varies. Error bars are computed over $10$ independent runs. The vertical red line marks the instability threshold $\gamma_{\rm stab}^2=0.0152$. Even in the unstable phase, the convergence criterion is well satisfied $\langle F \rangle_1-\langle F \rangle_2=0$.}
    \label{fig:RS_conv}
\end{figure*}

\paragraph{Correlated disorder and the cavity method}
Within the statistical-physics framework, this inference problem can be formally seen as a spin-glass model, where the probability distribution (\ref{eq:posterior}) is defined over a set of {\it dynamical} (or annealed) variables, which are the inferred infection times $\{t_i\}_{i\in V}$. Additionally, the distribution parametrically depends on a set of {\it disordered} (quenched) variables, namely the set of observations $\mathcal{O}=\{o_i\}_{i\in V}$.
These variables depend on the individuals' state, after some time has passed since the beginning of the dynamical process.
It is therefore clear that non-local correlations between the disordered variables $\mathcal{O}=\{o_m\}$ arise due the past history of the epidemic trajectory.

Studying spin-glass models with correlated disorder with classical statistical physics tools such as the cavity method is a challenge.
Indeed, a direct use of the cavity method needs a probability measure whose underlying graph of interaction (factor graph) is locally tree-like, and independent disordered variables.
In \cite{BrBuMa23} we proposed a strategy to overcome this difficulty, and used the cavity method under the Replica Symmetric ansatz to provide quantitative predictions on the properties of the posterior distribution (\ref{eq:posterior}) averaged over random ensembles of instances, in the thermodynamic limit.

\paragraph{Replica Symmetric cavity method}
There are different versions of the cavity method, that rely on self-consistent hypothesis on the effect of the long loops. 
The simplest version, called Replica Symmetric (RS), assumes a fast decay of the correlations between distant variables, in such a way that the measure will be correctly described by the locally tree-like approximation, and that the BP equations converge to a unique fixed-point on a typical large instance.
We recall in appendix~\ref{app:RS} the derivation made in \cite{BrBuMa23} of RS cavity equations (equations (\ref{eq:RS_eqn})).

\paragraph{Instability of the RS solution}
When the hypothesis underlying the RS cavity method breaks down, a more sophisticated version of the cavity method can be employed.
The first non-trivial level is called 1RSB (one step of RSB), and takes into account the effect of long loops in the factor graph.
The 1RSB formalism postulates the existence of a partition of the configuration space into pure states (or clusters) such that the restriction of the measure to a cluster is accurately described within the RS formalism.
Technically, a 1RSB phase can be described within the 1RSB formalism by deriving and solving the 1RSB equations (cf 1RSB equations in appendix~\ref{app:RS_stab}).
In the RS phase, the 1RSB cavity equations admit a unique trivial solution describing the RS phase.
An RSB phase transition is then unveiled by the appearance of a non-trivial solution of the 1RSB equation.
One way to investigate the existence of a non-trivial solution is to study the local stability of the RS solution, under a perturbation towards the space of 1RSB solutions. This is the approach adopted here to unveil an RSB phase transition, see appendix~\ref{app:RS_stab}.

\section{Results}

\subsection{RSB Phase transition}
\label{subsec:instability_RS}

\paragraph{Convergence towards an RS fixed-point}
The RS equations (\ref{eq:RS_eqn}) can be solved numerically with an iterative procedure, called population dynamics \cite{MePa00}. 
In this paragraph, we show that this iterative procedure always reaches a stationary regime (i.e., an RS fixed-point) in the range of parameters studied, even in the unstable phase.
Nonetheless, in the next paragraph we show that this RS fixed-point is unstable towards an RSB solution for $\gamma \in [\gamma_{\rm stab}^1,\gamma_{\rm stab}^2]$ (see Fig.~\ref{fig:phase_diagram}).

The left panel of Fig.~\ref{fig:RS_conv} shows the evolution under iterations of the free-energy for two different values of the seed probability $\gamma$.
The red upper curve corresponds to $\gamma=0.017>\gamma_{\rm stab}^2=0.0152$, i.e.\ to the phase where RS fixed-point is stable under perturbation towards the RSB function space, while the blue lower curve at $\gamma=0.015<\gamma_{\rm stab}^2$ is in the unstable phase (see Fig.\ref{fig:phase_diagram}).
Both curves fluctuate around a mean value, and the fluctuations can be interpreted as a finite-size effect of the population used to represent the RS distribution of BP messages. 
Despite the fluctuations, an estimate for the free-energy can be computed from the mean value over iterations in both cases.
Therefore, from the sole observation of the evolution of average quantities such as the free-energy, one cannot observe the instability transition appearing for $\gamma<\gamma_{\rm stab}^2$, as RS iterations still converge to a fixed-point even in the unstable phase. 

The right panel of Fig.~\ref{fig:RS_conv} displays a convergence criterion computed from the free-energy, for various values of $\gamma$ inside and outside of the unstable phase. 
The convergence criterion is computed as follows: for a given run of the iterative algorithm solving the RS equations, one computes the free-energy averaged over two time windows (see left panel of Fig.~\ref{fig:RS_conv}). 
The difference between the two resulting averages $\langle F \rangle_1-\langle F \rangle_2$ gives a criterion for estimating the convergence toward a fixed-point. 
In the present case, this criterion is always satisfied as $\langle F \rangle_1-\langle F \rangle_2$ is always statistically compatible with $0$ (see right panel of Fig.\ref{fig:RS_conv}), even in the unstable phase $\gamma<\gamma_{\rm stab}^1=0.0152$ (the error bars are computed over $10$ independent runs).

\paragraph{Stability criterion}

In this paragraph, we study the instability of the RS fixed-point towards the space of RSB solutions. 
This is done as follows: after reaching the RS fixed-point ($250$ iterations in our numerical experiments), we slightly perturb the system.
The variance of the perturbation, averaged over the population is denoted $\Delta(t)$, and its expression and update-rule are given in appendix~\ref{app:RS_stab}, equations (\ref{eq:Delta_M}, and \ref{eq:1st_mom}).

\begin{figure}
	\centering
	\includegraphics[width=\columnwidth]{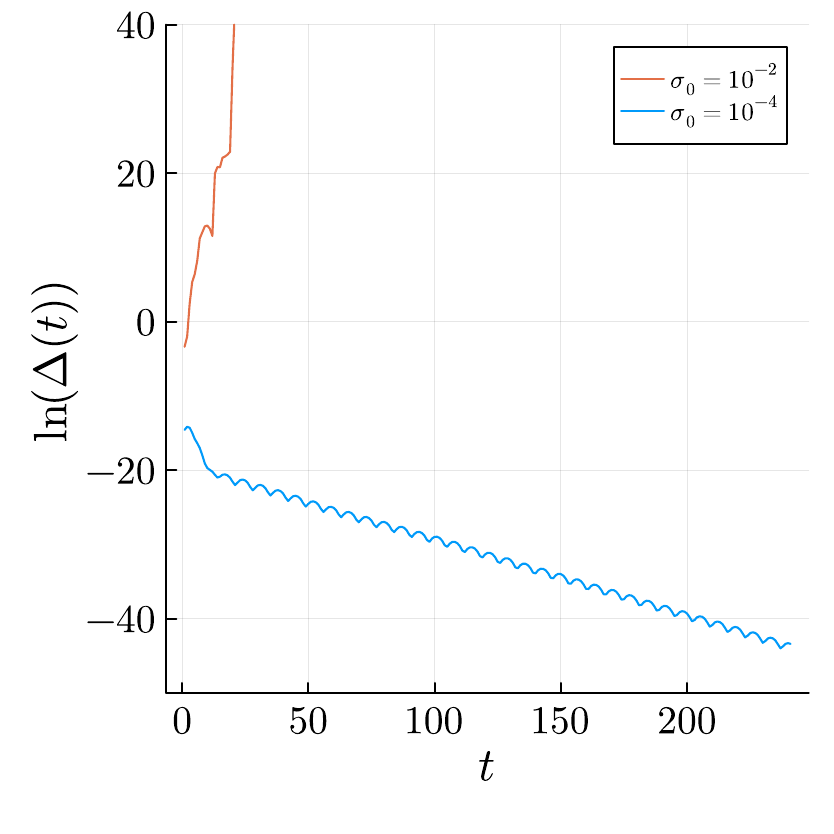}
	\caption{
		Evolution of the stability parameter $\Delta(t)$ under iterations, for seed probability $\gamma=0.016$ and two different initial conditions $\sigma_0$ (see main text and appendix \ref{appsubsec:stab_IC}). The transmission rate is $\lambda=1$, observation time $T_\text{obs}=8$, and population-size $\mathcal{N}=5\cdot10^4$.}
	\label{fig:RS_instab_twoIC}
\end{figure}

We show in Fig.~\ref{fig:RS_instab_twoIC} the evolution of the stability parameter $\Delta(t)$ under iterations, for a given value of the seed probability $\gamma=0.016$, and two initial conditions for the perturbation $\Delta(t=0)$. At time $t=0$, the perturbation in each message is set to $M_i(T)=\sigma_0\times\mu_i(T)$, for each element $i\in\{1,\dots,\mathcal{N}\}$ of the population representing the RS distribution, and for each value of the variable $T$ (see appendix \ref{appsubsec:stab_IC}).
We see that the evolution of the stability parameter depends strongly on the initial condition: the one with a smaller initial variance leads to an exponential decay of the stability parameter, with an exponential rate that can be estimated with a linear fit of $\ln(\Delta(t))$.
Conversely, the initial condition with a larger variance leads to a divergence of the stability parameter. 
This can be understood as follows: even in a regime where the RS solution is stable (we are in the stable phase at $\gamma>\gamma_{\rm stab}^2$), a strong perturbation might be able to destabilize the fixed-point found numerically with a {\it finite-size} population $\mathcal{N}$. 

For a small enough initialization $\Delta(0)$ of the perturbation, one can extract the decay rate $\delta$, defined by 
\begin{align}
\Delta(t)\simeq \delta^t\;,
\end{align}
through a simple linear fit to $\ln(\Delta(t))$.
The instability of the RS fixed-point towards RSB can be then located at the $\gamma$ value where the rate $\delta$ becomes greater than 1.

\begin{figure*}
	\centering
	\includegraphics[width=0.4\textwidth]{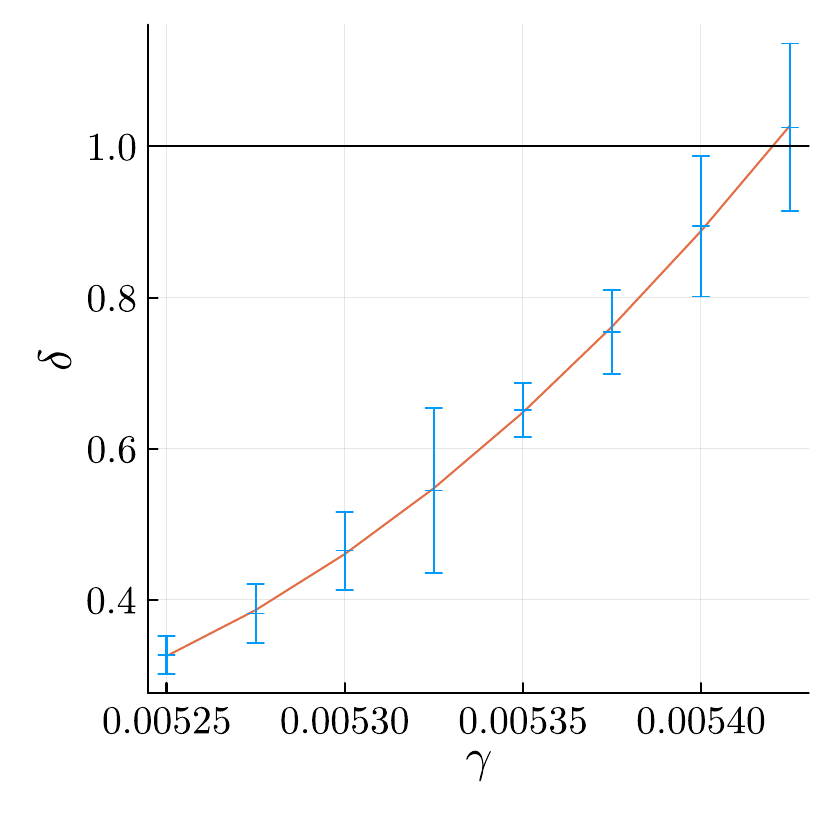} \
	\includegraphics[width=0.4\textwidth]{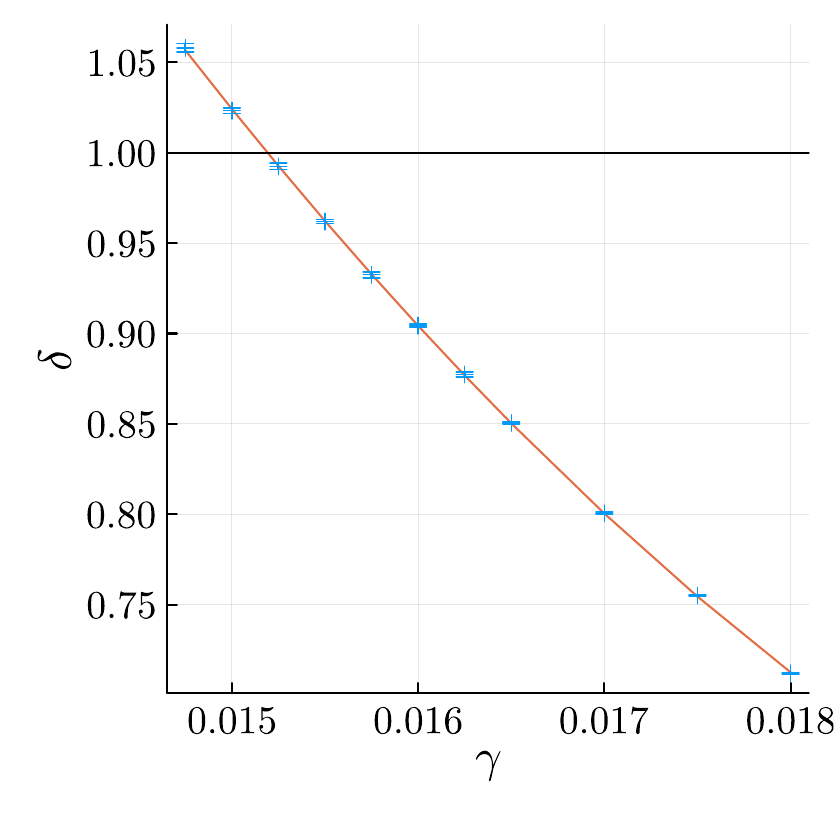}
	\caption{Stability study of the RS fixed-point under a perturbation towards the space of RSB solutions. Left: stability parameter $\delta$ approaching the value $1$ for $\gamma\nearrow\gamma_{\rm stab}^1=0.00542$. Right: stability parameter $\delta$ approaching the value $1$ for $\gamma\searrow\gamma_{\rm stab}^2=0.0152$. Error bars are computed over $10$ independent runs, population size is set to $\mathcal{N}=5\cdot10^5$ for the left plot, $5\cdot10^4$ for the right plot. Red curves show a quadratic fit giving the threshold values.}
	\label{fig:RS_instab_fit}
\end{figure*}

Figure \ref{fig:RS_instab_fit} shows the stability parameter $\delta$ as the probability seed $\gamma$ varies, for transmission rate $\lambda=1$, observation time $T_\text{obs}=8$, on random regular graphs with degree $d=3$.
One can clearly identify two stability thresholds $\gamma_{\rm stab}^1< \gamma_{\rm stab}^2$ (obtained from a quadratic fit) such that:
\begin{itemize}
	\item the stability parameter $\delta$ grows towards the value $1$ for $\gamma\nearrow\gamma_{\rm stab}^1=0.00542$
	\item the stability parameter $\delta$ grows towards the value $1$ for $\gamma\searrow\gamma_{\rm stab}^2=0.0152$
\end{itemize}
We have therefore identified an RSB phase for $\gamma\in[\gamma_{\rm stab}^1,\gamma_{\rm stav}^2]$ for which the RS fixed-point is unstable under a perturbation towards the RSB space.

\subsection{Instability of Belief Propagation on large graphs instances}
\label{subsec:instability_BP}

\begin{figure}
	\centering
	\includegraphics[width=0.8\columnwidth]{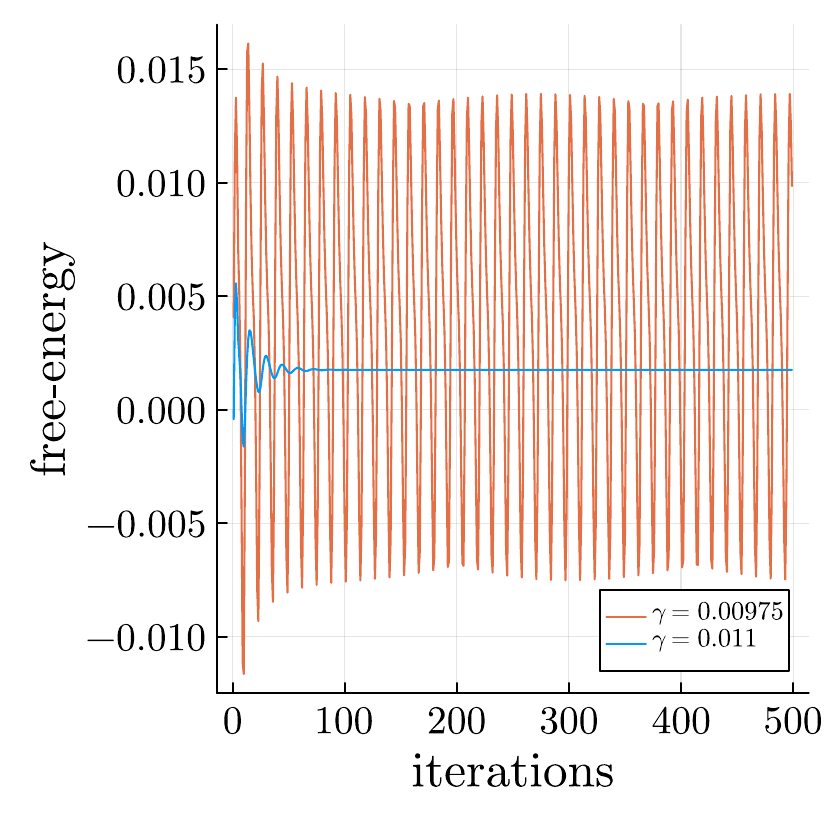}
	\caption{Evolution of the free-energy along iterations for the resolution of the BP equations on a finite-size instance ($N=5\cdot10^5$), for two different values of the seed probability $\gamma$ (for transmission rate $\lambda=1$, observation time $T_\text{obs}=8$, on a $3$-regular random graph). For $\gamma=0.011>\gamma_{\rm BP}^2(N)=0.010$, BP reaches a fixed-point and the free-energy converges, while for $\gamma=0.00975<\gamma_{\rm BP}^2(N)$ BP fails at finding a fixed-point and the free-energy oscillates.}
	\label{fig:BP_iterations}
\end{figure}

In this section, we analyze the stability of the fixed-point of the BP equations (\ref{eq:BP_posterior}) in finite-size instances of the inference problem, where an instance is defined by a contact graph $G=(V,E)$ and a realization of the individuals' trajectories $\{t_i\}_{i\in V}$ and observations $\mathcal{O}$.
The BP equations do not converge for a range of $\gamma$ values close to $\gamma\simeq 0.01$, for a high transmission rate $\lambda\simeq 1$ (see  \cite{GhArBiZd23}). Figure~\ref{fig:BP_iterations} illustrates this lack of convergence for $\gamma=0.00975$ (red curve) of the BP updates. 

\begin{figure*}
	\centering
	\includegraphics[width=0.45\textwidth]{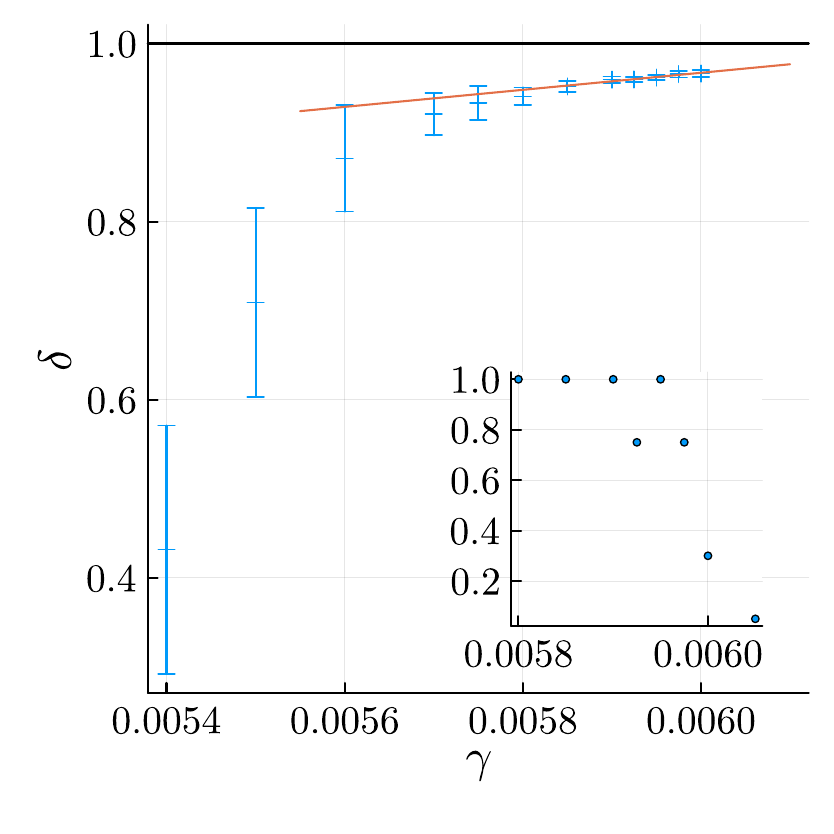}
	\includegraphics[width=0.45\textwidth]{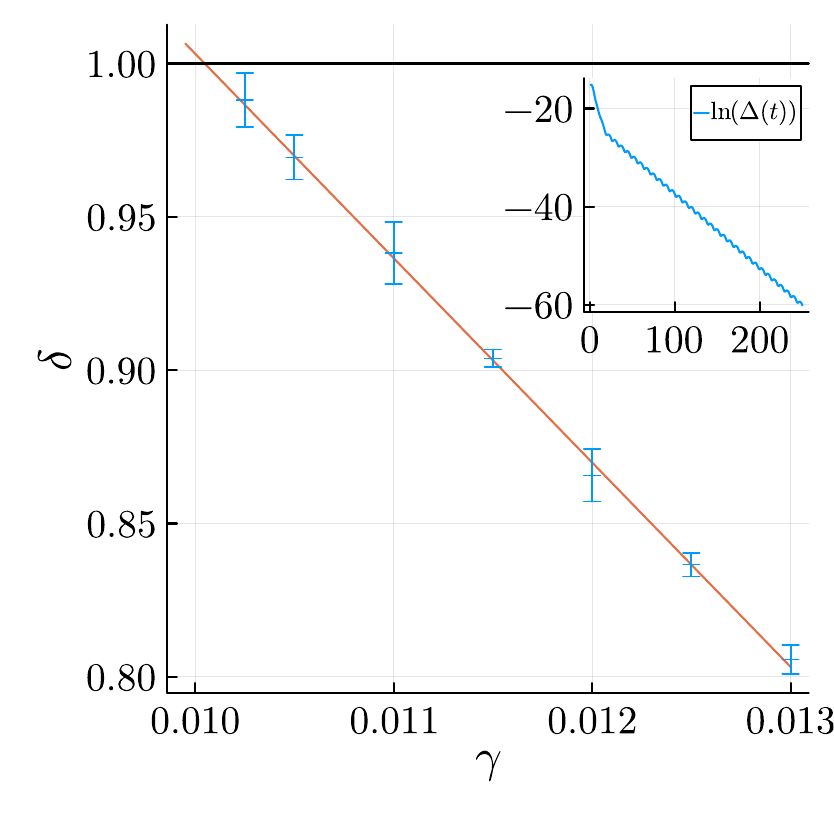}
	\caption{Stability of the BP fixed-point equation on $3$-regular random graphs with $\lambda=1$ and $T_\text{obs}=8$.  Main panels: the stability parameter $\delta(G,\mathcal{O})$ defined in Eq.~(\ref{eq:stab_1instance}) approaches the value $1$, for $\gamma\nearrow\gamma_{\rm BP}^2(N)\simeq0.0063$ (left) and for $\gamma\searrow\gamma_{\rm BP}^1(N)=0.010$ (right). Results are averaged over 20 instances of size $N=10^6$ (left) and 10 instances of size $N=5\cdot10^5$ (right). The red lines are linear fits. Inset in left panel: the fraction of BP converged runs goes to $0$ as $\gamma$ approaches the value $\gamma_{\rm BP}^2(N)$. Inset in the right panel: the evolution of the perturbations variance $\ln\Delta(t;G,\mathcal{O})$ along iterations, for $\gamma=0.012$, exhibits an exponential decay.}
	\label{fig:BP_instab_fit}
\end{figure*}

The detailed protocol for the stability analysis is given in appendix~\ref{app:BP_stab}, and summarized here.
On a single instance, after having reached a fixed-point (assuming that we are in a regime where we are able to find one), we switch on a perturbation $\epsilon_{i\to j}$ for each BP message $\mu_{i\to j}$ on each directed edges of the contact graph. 
We then track the evolution of the perturbation under the iterations (\ref{eq:stab_onegraph_update}). The magnitude of the perturbation is measured by computing:
\begin{align}
\label{eq:stab_onegraph}
	\Delta(t;G,\mathcal{O}) = \frac{1}{2|E|}\sum_{i\in V}\sum_{j\in\partial i}\sqrt{\frac{1}{q}\sum_{T\in\chi}\epsilon_{i\to j}(T)^2}
\end{align}
We show in the inset in the right panel of Fig.~\ref{fig:BP_instab_fit} the evolution of the quantity $\Delta(t;G,\mathcal{O})$ for a given instance of size $N=5\cdot10^5$. We observe again an exponential decay of the magnitude of the perturbations, with a decay rate $\delta(G,\mathcal{O})$ defined by
\begin{align}
\label{eq:stab_1instance}
	\Delta(t;G,\mathcal{O}) = \delta(G,\mathcal{O})^t
\end{align}
and that can be easily extracted from a linear fit to $\ln \Delta(t;G,\mathcal{O})$.
The main panels of Fig.~\ref{fig:BP_instab_fit} display this decay rate for various values of $\gamma$. 
Note that in order to avoid fluctuations in the number of seeds in a finite-size instance, we fixed the number of seeds to be exactly equal to $\gamma N$, rather than fixing only the average number as in the previous section. 
We can again identify (size-dependent) stability thresholds $\gamma_{\rm BP}^2(N)=0.010$ (see the linear fit in the right panel of Fig.~\ref{fig:BP_instab_fit}) such that the stability parameter (decay rate) $\delta$ grows toward the value $1$ for $\gamma\searrow\gamma_{\rm BP}^2(N)$.

The stability threshold $\gamma_{\rm BP}^1(N)$ is harder to identify from the stability analysis, as shown in the left panel of Fig.~\ref{fig:BP_instab_fit}. The stability parameter $\delta(G,\mathcal{O})$ jumps from small values (around $0.4$) to values approaching $1$ in a tiny interval ($0.0054\lesssim\gamma\lesssim0.006$).
At the same time, the fraction of converged BP runs drops to zero around $\gamma\simeq0.006$ (see the inset in the left panel).
A threshold value can still be estimated from the linear fit in the left main panel, leading to $\gamma^1_{\rm BP}(N)\simeq 0.0063$.

Note that the interpretation of this stability study is slightly different from the analysis performed in the previous section \ref{subsec:instability_RS}. 
In the analysis of the RS solution via the population dynamics, we were able to find a fixed-point even in the unstable RSB phase, and thus the instability threshold can be estimated from both sides, providing a precise location of the RSB phase boundary.
Instead, BP on finite-size instances fails to find a fixed-point in the unstable phase (see Fig.~\ref{fig:BP_iterations}), and thus the stability analysis can be performed only coming from the stable phase, giving a less accurate location of the RSB phase.

Finally, we can observe that the unstable region for BP $[\gamma_{\rm BP}^1(N),\gamma_{\rm BP}^2(N)]=[0.0063, 0.010]$ is narrower than the RSB phase: $[\gamma_{\rm stab}^1,\gamma_{\rm stab}^2]=[0.00542, 0.0152]$. 
The fact that BP can still converge towards a fixed-point, even in an RSB phase, can be interpreted as a sign of a continuous RSB transition, where correlations between distant variables due to the presence of long loops arise progressively when entering the RSB phase.
Note however that a confirmation that the observed RSB transition is continuous would require the full resolution of the 1RSB equations (see appendix \ref{app:RS_stab}, equation (\ref{eq:1rsb_av})) to exclude the existence of a competing discontinuous phase transition towards a 1RSB solution.

\section{Discussion and Perspectives}

In this paper, we unveil a RSB phase transition in the epidemic inference problem, under the Nishimori conditions (Bayes optimal setting).
We identify the RSB thresholds through a stability study of the RS solution under a perturbation towards the space of RSB solutions.

The stability study done in this paper only allows us to unveil a continuous RSB phase transition, where the RS solution becomes unstable.
In principle, a non-trivial fixed-point to the full 1RSB equations can appear discontinuously, even in the phase where the RS solution is stable (i.e. for $\gamma<\gamma_{\rm stab}^1$, or for $\gamma>\gamma_{\rm stab}^2$, see Fig.~\ref{fig:phase_diagram}).
However, for the aim of this work --- i.e., showing the presence of RSB under the Nishimori conditions in epidemic models --- the existence of a continuous transition towards RSB is enough. The eventual presence of a further discontinuous phase transition could make the scenario even more different from the previous common belief that the Nishimori condition implies an RS solution.
We leave the study of the full 1RSB equations and the search for a discontinuous 1RSB phase transition to future works.

The main result of this work naturally poses the question of why the Nishimori conditions do not imply an RS solution.
A possible answer comes from the observation that in these epidemic models the disorder variables (i.e., the observations) are correlated.
In this context, we leave for future work the study of correlated disorder on simpler models than in the epidemic framework, starting with the canonical planted Ising model.
Very recently Nishimori pointed out \cite{Nishimori24,Nishimori25} that, in spin glass models with correlated disorder, RSB effects can appear also under the Nishimori conditions.

This is a crucial observation for practical applications.
Indeed, in real-world datasets, one can not assume randomness enters in an uncorrelated way and so the use of the inference methods based on the RS assumption is at odds.
Well-known inference algorithms like AMP \cite{ZdKr16} or BP \cite{RiSeZd19} may dramatically fail under these realistic conditions.
For this reason, it is crucial to better understand the limits of more robust algorithms like those based on Monte Carlo Markov Chain in solving inference problems \cite{angelini2023limits}.

\acknowledgments
LB acknowledges discussions with Ahmed El Alaoui, Guilhem Semerjian and Lenka Zdeborov\'a.

The research has been supported by the first FIS (Italian Science Fund) 2021 funding scheme (FIS783 - SMaC - Statistical Mechanics and Complexity) from MUR, Italian Ministry of University and Research and by the “National Centre for HPC, Big Data and Quantum Computing - HPC”, Project CN\_00000013, CUP B83C22002940006, NRP Mission 4 Component 2 Investment 1.5, Funded by the European Union - NextGenerationEU.

This study was carried out within the FAIR - Future Artificial Intelligence Research project and received funding from the European Union NextGenerationEU (Piano Nazionale di Ripresa e Resilienza (PNRR)–Missione 4 Componente 2, Investimento 1.3–D.D. 1555 11/10/2022, PE00000013). This manuscript reflects only the authors’ views and opinions, neither the European Union nor the European Commission can be considered responsible for them.

\bibliography{biblio.bib}

\appendix

\section{Replica Symmetric formalism for Epidemic Inference}
\label{app:RS}

In this section, we recall the Replica Symmetric formalism derived in \cite{BrBuMa23} to study the posterior probability (\ref{eq:posterior}).

\subsection{Correlated observations}

We denote by $\mathcal{D}=\{\{x_i^0\}_{i\in V},\{s_{ij},s_{ji}\}_{(i,j)\in E}\}$ the set of disordered variables, and let $\mathcal{D}_i=\{x_i^0, \{s_{li}\}_{l\in\partial i}\}$ be the local disordered variables at node $i\in V$.

The joint probability on the planted times $\{\tau_i\}_{i\in V}$, on the observations $\mathcal{O}$ and on the inferred times $\{t_i\}_{i\in V}$, conditioned on the disorder $\mathcal{D}$ can be decomposed using chain rule:
\begin{align}
P(\underline{t},\mathcal{O},\underline\tau|\mathcal{D}) &= P(\underline\tau|\mathcal{D})P(\mathcal{O}|\mathcal{D},\underline\tau)P(\underline{t}|\mathcal{D},\underline\tau,\mathcal{O})
\label{eq:intro:joint_total_distribution}
\end{align}
Where each term is given as follows:
\begin{enumerate}
	\item The first term corresponds to sampling the planted times conditioned on the disorder $\mathcal{D}$ from the prior, cf equation (\ref{eq:prior:constraints_delay_times}):
	$$
	P(\tau|\mathcal{D}) = \prod_{i\in V}\psi^*(\tau_i,\underline{\tau}_{\partial i};\mathcal{D}_i) 
	$$
	\item The second term is sampling the observations $\mathcal{O}$, conditioned on the planted times and on the disorder, according to equation (\ref{eq:observations}): 
	$$
	P(\mathcal{O}|\mathcal{D},\underline\tau) = \prod_{i\in V}\rho(o_i,\tau_i)
	$$
	\item Finally, the last term $P(\underline{t}|\mathcal{O},\mathcal{D},\underline\tau)$ is independent of the disordered variables $\mathcal{D}$ and the planted times $\underline{\tau}$, since the inferred times are sampled from the posterior probability distribution:
	$$
	P(\underline{t}|\mathcal{D},\underline\tau,\mathcal{O}) = \frac{P(\underline{t})P(\mathcal{O}|\underline{t})}{P(\mathcal{O})}
	$$
	with $P(\mathcal{O}|\underline{t})$ given by equation (\ref{eq:observations}), and with:
	\begin{align*}
	P(\underline{t}) &= \prod_{i\in V}\psi(t_i,\underline{t}_{\partial i}) \quad \text{where:} \\
	\psi(t_i,\underline{t}_{\partial i}) &=\sum_{\mathcal{D}_i}\psi^*(t_i,\underline{t}_{\partial i};\mathcal{D}_i) 
	\end{align*}
\end{enumerate}
Note that for a given realization of the disorder $\mathcal{D}$, the planted times $\underline{\tau}$ are fixed, and so are the (noiseless) observations $\mathcal{O}$. We hence denote the denominator $P(\mathcal{O})$ as a function of the disordered variables $\mathcal{D}$:
$$
P(\mathcal{O})=Z(\mathcal{D})
$$
By summing over the observations (the only non-zero term for $\mathcal{O}=\mathcal{O}(\mathcal{D})$), we obtain finally an expression for the joint probability distribution over the planted and the inferred times, that factorizes over the sites of the contact graph $G=(V,E)$:
\begin{align}
\label{eq:joint_short_loops}
\begin{aligned}
	&P(\underline{\tau},\underline{t}|\mathcal{D}) = \frac{1}{Z(\mathcal{D})}\prod_{i\in V}\psi^*(\tau_i,\underline{\tau}_{\partial i};\mathcal{D}_i)\psi(t_i,\underline{t}_{\partial i})\xi(t_i,\tau_i) \\
	&\text{with} \quad \xi(t_i,\tau_i)=\delta_{\tau_i\leq T_{\rm obs}}\delta_{t_i\leq T_{\rm obs}} + \delta_{\tau_i> T_{\rm obs}}\delta_{t_i> T_{\rm obs}}
\end{aligned}
\end{align}

\subsection{Auxiliary variables}
\label{appsubsec:aux_vars_joint}
The factor graph associated with the probability distribution (\ref{eq:joint_short_loops}) contains short loops which prevent from a direct use of the cavity method.
To remove these short loops, we introduce the auxiliary variables $T_{ij} = \{\tau_i^{(j)},\tau_j^{(i)},t_i^{(j)},t_j^{(i)}\}$ defined on the edges of the contact graph $G=(V,E)$, which are copies of the inferred and planted times:
$$
\tau_i^{(j)}=\tau_i \quad t_i^{(j)}=t_i \quad \forall i\in V, j\in \partial i
$$
We obtain the joint probability distribution over the auxiliary variables  $T_{ij}=(\tau_i^{(j)},\tau_j^{(i)},t_i^{(j)},t_j^{(i)})$, $(i,j)\in E$:
\begin{align}
	\label{eq:prob_auxiliary}
	P(\{T_{ij}\}_{(ij)\in E}|\mathcal{D}) &= \frac{1}{\mathcal{Z}(\mathcal{D})}\prod_{i\in V}\Psi(\{T_{il}\}_{l\in\partial i};\mathcal{D}_i) 
\end{align}
and with (for any $j\in \partial i$):
\begin{align}
	\begin{aligned}
		\Psi(\{T_{il}\}_{l\in\partial i};&\mathcal{D}_i) = \xi(\tau_i^{(j)},t_i^{(j)})\\
		\times &\psi^*(\tau_i^{(j)},\underline{\tau}_{\partial i}^{(i)};\{s_{li}\}_{l\in\partial i},x_i^0)\\
		\times&\psi(t_i^{(j)},\underline{t}_{\partial i}^{(i)})\prod_{l\in\partial i\setminus j}\delta_{t_i^{(j)},t_i^{(l)}}\delta_{\tau_i^{(j)},\tau_i^{(l)}}
	\end{aligned}
\end{align}

\subsection{BP equations for the joint probability}
\begin{figure*}
	\centering
	\includegraphics[width=0.8\textwidth]{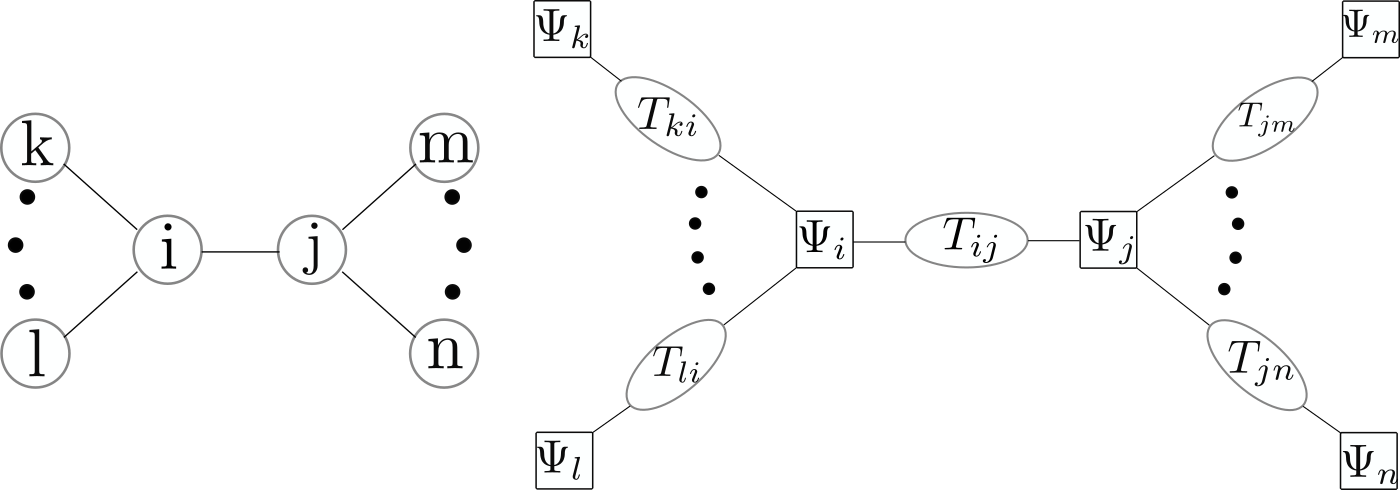} 	
	\caption{Left: contact graph $G=(V,E)$. Right: factor graph for the probability (\ref{eq:prob_auxiliary}): the auxiliary variables $T_{ij}$ live on the original edges $(i,j)\in E$ of the contact graph $G$, and the factor functions $\Psi_i$ live on the original vertices $i\in V$ of the contact graph $G$.}
	\label{fig:factor_graph}
\end{figure*}
On a single instance (given by a contact graph $G=(V,E)$ and a realization of the disorder $\mathcal{D}$), the factor graph associated with the probability distribution (\ref{eq:prob_auxiliary}) is given in Figure~\ref{fig:factor_graph} (right panel).
By introducing the auxiliary variables $\{T_{ij}\}_{(i,j)\in E}$, we have removed the systematic short loops arising in (\ref{eq:joint_short_loops}). 
The nodes representing the auxiliary variables $T_{ij}$ live on the edges of the original contact graph, and the factor functions $\Psi_i, i\in V$ live on its vertices.
The resulting factor graph has the same structure as the original contact graph, in particular, if $G$ is a tree, then the following BP equations are exact.
For each directed edges of the contact graph $G=(V,E)$, we define the variable-to-factor BP message $\mu_{i\to \Psi_j}(T_{ij})$, going from variable node $(i,j)$ to factor node $\Psi_j$.
Similarly, we define the factor-to-variable $\nu_{\Psi_i\to j}(T_{ij})$, going from factor node $\Psi_i$ to variable node $(i,j)$. 
Since the variable $(i,j)$ node representing $T_{ij}$ is of degree $2$, we have the trivial equality:
$$
\mu_{i\to \Psi_j}(T_{ij})=\nu_{\Psi_i\to j}(T_{ij})
$$ 
The BP message $\mu_{i\to \Psi_j}(T_{ij})$ is the marginal probability of the variable $T_{ij}$ in the graph $G$ amputated from node $j$.

The BP equations for the joint probability distribution, conditioned on the disorder $\mathcal{D}$ can be written:
\begin{align}
\label{eq:BP_equations}
\begin{aligned}
\mu_{i\to \Psi_j}(T_{ij}) &= \\ 
\frac{1}{z_{\Psi_i\to j}}&\sum_{\{T_{il}\}_{k\in\partial i \setminus j}}\Psi(\{T_{il}\}_{l\in\partial i};\mathcal{D}_i)\prod_{k\in\partial i \setminus j}\mu_{k\to\Psi_i}(T_{ik})
\end{aligned}
\end{align}	
were $z_{\Psi_i\to j}$ is a normalization factor. 

The free-energy $F=-\log(Z(\mathcal{D})=-\log(P(\mathcal{O})$ can be expressed in term of the BP-messages:
\begin{align}
F(G,\mathcal{D}) = \frac{1}{N}\sum_{i\in V}\left(\frac{d}{2}-1\right)\ln Z_{\Psi_i} - \frac{1}{2N}\sum_{i\in V}\sum_{j\in\partial i}\ln z_{\Psi_i\to j}
\end{align}
with
\begin{align}
	Z_{\Psi_i} = \sum_{\{T_{ij}\}_{j\in\partial i}} \Psi(\{T_{il}\}_{l\in\partial i};\mathcal{D}_i)\prod_{l\in\partial i}\mu_{l\to\Psi_i}(T_{il})
\end{align}

\subsection{Simplifications for the BP equations}
\label{appsubsec:simplifs_BP_joint}
As already noted in \cite{BrBuMa23}, the variable $T_{ij}$ is a variable made of $4$ infection times.
In the numerical resolution of the BP equations, it will be convenient to introduce a horizon time $\mathcal{T}$, above which the epidemic evolution is not observed. 
The size of variable $T_{ij}$ is then $(T+1)^4$, but it is possible to decrease the size of the variable on which the BP message is defined.
By inspecting the BP equation, we note that the r.h.s. depends on the planted time $\tau_j^{(i)}$ only through the sign $\sigma_{ji}\in\{-1,0,1\}$ defined as:
\begin{align}
	\sigma_{ji} = 1+{\rm sgn}(\tau_j^{(i)}-\tau_i^{(j)}+s_{ji})
\end{align}
We therefore introduce the following auxiliary messages:
\begin{align}
 \tilde{\nu}_{\Psi_i\to j}(\tau_i^{(j)},\sigma_{ji},t_i^{j},t_j^{i}) = \nu_{\Psi_i\to j}(\tau_i^{(j)},\tau_j^{(i)},t_i^{(j)},t_j^{(i)})
\end{align}
for all $\tau_j^{(i)}$ such that $\sigma_{ji} = 1+{\rm sgn}(\tau_j^{(i)}-\tau_i^{(j)}+s_{ji})$.
We also introduce the auxiliary message:
\begin{widetext}
\begin{align}
\tilde{\mu}_{i\to\Psi_j}(\sigma_{ij},\tau_j^{(i)},c_{ij},t_j^{(i)}) = \sum_{t_i^{(j)}}a(t_j^{(i)}-t_j^{(i)}-c_{ij})\sum_{\tau_i^{(j)}}\mathbb{I}[\sigma_{ij}=1+{\rm sgn}(\tau_i^{(j)}-\tau_j^{(i)}+s_{ij})]\mu_{i\to\Psi_j}(\tau_i^{(j)},\tau_j^{(i)},t_i^{(j)},t_j^{(i)})
\end{align}
\end{widetext}
with $c_{ij}\in\{0,1\}$ and $a(t)=(1-\lambda)^{H(t)}$, and $H(t)$ the Heaviside function.
We can obtain a set of closed equation on these auxiliary messages (for details on the derivation, see \cite{BrBuMa23}, appendix A.):
\begin{widetext}
	\begin{align}
	\label{eq:BP_equations_simplif}
	\begin{aligned}
	\tilde{\mu}_{i\to \Psi_j}(\sigma_{ij},\tau_j^{(i)},c_{ij},t_j^{(i)}) &= \sum_{t_i^{(j)}}a(t_j^{(i)}-t_j^{(i)}-c_{ij})\sum_{\tau_i^{(j)}}\tilde{\nu}_{\Psi_i\to j}(\tau_i^{(j)},\sigma_{ji}=1+\text{sgn}(\tau_j^{(i)}-\tau_i^{j}+s_{ji}),t_i^{(j)},t_j^{(i)})\\
	&\times\mathbb{I}[\sigma_{ij}=1+\text{sgn}(\tau_i^{(j)}-\tau_j^{(i)}+s_{ij})] \\
	\text{and}&\\
	\tilde{\nu}_{\Psi_i\to j}(\tau_i^{(j)},\sigma_{ji},t_i^{(j)},t_j^{(i)})&=\gamma(t_i^{(j)})\xi(\tau_i^{(j)},t_i^{(j)})(A_1(\tau_i^{(j)},\sigma_{ji},t_i^{(j)},t_j^{(i)})-\phi(t_i^{(j)})A_0(\tau_i^{(j)},\sigma_{ji},t_i^{(j)},t_j^{(i)}))  \\
	\end{aligned}
	\end{align}
with: 
\begin{align}
	\begin{aligned}
	A_c(\tau_i^{(j)},\sigma_{ji},t_i^{(j)},t_j^{(i)})&=	a(t_i^{(j)}-t_j^{(i)}-c)\delta_{x_i^0,I}\delta_{\tau_i^{(j)},0}\prod_{k\in\partial i\setminus j}\left(\sum_{\sigma_{ki}=0}^2\tilde{\mu}_{k\to\Psi_i}(\sigma_{ki},\tau_i^{(k)},c_{ki}=c,t_i^{(k)})\right) \\
	&+a(t_i^{(j)}-t_j^{(i)}-c)\delta_{x_i^0,S}\delta_{\sigma_{ij}\in\{1,2\}}\prod_{k\in\partial i\setminus j}\left(\sum_{\sigma_{ki}=1}^2\tilde{\mu}_{k\to\Psi_i}(\sigma_{ki},\tau_i^{(k)},c_{ki}=c,t_i^{(k)})\right) \\
	&-a(t_i^{(j)}-t_j^{(i)}-c)\delta_{x_i^0,S}\delta_{\tau_i^{(j)}<\mathcal{T}}\delta_{\sigma_{ij}=2}\prod_{k\in\partial i\setminus j}\left(\tilde{\mu}_{k\to\Psi_i}(\sigma_{ki}=2,\tau_i^{(k)},c_{ki}=c,t_i^{(k)})\right) \\
	\end{aligned}
\end{align}
and 
$$
\gamma(t_i)=\gamma\delta_{t,0}+(1-\gamma)\delta_{t>0}
$$
\end{widetext}
\subsection{Replica Symmetric cavity equations}
In order to describe the properties of (\ref{eq:prob_auxiliary}) averaged over the random ensemble of instances, we now apply the cavity method under the Replica Symmetric (RS) hypothesis.
We remind that an instance is defined by a contact graph $G=(V,E)$ (we consider here random $d$-regular random graphs), and a realization of the disordered variables $\mathcal{D}$.
In the RS formalism, one assumes that the effect of long loops is negligible, and that the distribution (\ref{eq:prob_auxiliary}) is well described by the unique fixed-point of the BP equations.
We consider a uniformly chosen directed edge $i\to j$ in a random contact graph, and let $\mathcal{P}^{\rm rs}(\mu)$ be the probability of the fixed-point message $\mu_{i\to \Psi_j}$ thus obtained. 
Then, under the RS hypothesis, the incoming messages on a given factor graph $\Psi_i$ are i.i.d. with distribution $\mathcal{P}^{\rm rs}$, implying that the probability distribution $\mathcal{P}^{\rm rs}$ must obey the following self-consistent equation:
\begin{align}
\begin{aligned}
\label{eq:RS_eqn}
	\mathcal{P}^{\rm RS}(\mu) &= \sum_{\mathcal{D}_{\rm s}}P(\mathcal{D}_{\rm s})\int\prod_{i=1}^{d-1}{\rm d}\mathcal{P}^{\rm RS}(\mu_i)\\
	&\times \delta[\mu-f(\mu_1,\dots,\mu_{d-1};\mathcal{D}_{\rm site})]
\end{aligned}
\end{align}
Where $\mathcal{D}_{\rm s}=\{x^0,s_1,\dots,s_d\}$ is a local disordered variable, and where $\mu-f(\mu_1,\dots,\mu_{d-1};\mathcal{D}_{\rm site})$ is a short-hand notation for the BP equation (\ref{eq:BP_equations}) (or (\ref{eq:BP_equations_simplif}), depending on the chosen representation of the BP messages).

The RS prediction for the free-energy is then:
\begin{align}
\label{eq:RS_free_energy}
\begin{aligned}
&F^{\rm RS} = - \frac{d}{2}\int{\rm d}\mathcal{P}^{\rm rs}(\mu)z_{\Psi_i\to j}(\mu) \\
&+\left(\frac{d}{2}-1\right)\sum_{\mathcal{D}_{\rm s}}P(\mathcal{D}_{\rm s})\int\prod_{i=1}^d{\rm d}\mathcal{P}^{\rm rs}(\mu_i) \ln Z_{\Psi_i}(\{\mu_i\}_{i=1\dots,d}) 
\end{aligned}
\end{align}

The RS equation (\ref{eq:RS_eqn}) can be solved numerically by population dynamics \cite{MePa00}, where the distribution $\mathcal{P}^{\rm RS}$ is approximated by the empirical distribution over a large sample of representative elements:
$$
\mathcal{P}^{\rm RS}(\mu)\simeq\frac{1}{\mathcal{N}}\sum_{i=1}^\mathcal{N}\delta(\mu-\mu_{i})
$$

\section{Instability of the RS solution}
\label{app:RS_stab}

\subsection{One-step RSB}
Under the 1RSB hypothesis, one assumes that the distribution (\ref{eq:prob_auxiliary}) is partitioned into clusters (or pure-states):
\begin{align}
P(\{T_{ij}\}|\mathcal{D})=\sum_{\gamma}p(\gamma)\mu_\gamma(\{T_{ij}\}|\mathcal{D})
\end{align}
with $p(\gamma)$ the distribution over the clusters.
The restriction of the joint distribution to one cluster $\mu_\gamma$ can be described by the RS formalism, i.e. can be described by a fixed-point of the BP equations.
We define $P_{i\to j}$ as the probability law of the message $\mu_{i\to j}^\gamma$, for a cluster $\gamma$ being chosen randomly with probability $p(\gamma)$.
Then the 1RSB messages $\{P_{i\to j}\}_{(i,j)\in E}$ obey the self-consistent equations:
\begin{widetext}
\begin{align}
\label{eq:1rsb_ongraph}
P_{i\to j}(\mu_{i\to j})=\frac{1}{Z_{i\to j}}\int\prod_{k\in\partial i\setminus j}{\rm d}P_{k\to i}(\mu_{k\to i})\delta[\mu_{i\to j}=f(\{\mu_{k\to i}\})_{k\in\partial i\setminus j}]z_{i\to j}(\{\mu_{k\to i}\})_{k\in\partial i\setminus j})^x
\end{align}
\end{widetext}
where $x$ is the Parisi parameter, allowing to weight differently the various clusters according to their size.
In order to average over the disorder, one introduces the probability distribution over the 1RSB messages: $\mathcal{P}^{\rm 1rsb}(P)$. 
It obeys the 1RSB equation (similar to the RS equation (\ref{eq:RS_eqn})):
\begin{align}
\label{eq:1rsb_av}
\mathcal{P}^{\rm 1rsb}(P)=\int\prod_{i=1}^{d-1} {\rm d}\mathcal{P}^{1rsb}(P_i)\delta[P-F(\{P_i\}_{i=1\dots,d})]
\end{align}
with $P=F(\{P_i\}_{i=1\dots,d})$ a shorthand notation for the equation (\ref{eq:1rsb_ongraph}).

\subsection{RS trivial fixed-point of the 1RSB equation:}
On a given graph, in the Replica Symmetric phase, there is only one fixed-point of the BP equation -- that we denote $\{\bar{\mu}_{i\to j}\}_{ij\in E}$ -- and the solution to the equation (\ref{eq:1rsb_ongraph}) is a Dirac delta:
\begin{align}
P_{i\to j}(\mu_{i\to j})=\delta(\mu_{i\to j},\bar{\mu}_{i\to j})
\end{align}
Once averaged over the disorder, we can see that the 1RSB equation (\ref{eq:1rsb_av}) always admit the following Replica Symmetric trivial solution:
\begin{align}
\label{eq:trivial_fp}
\mathcal{P}^{\rm 1rsb}(P)=\mathcal{P}^*(P)=\int{\rm d}\mathcal{P}^{\rm rs}(\bar{\mu})\delta[P(\mu)=\delta(\mu,\bar{\mu})]
\end{align}
In the RS phase, this is the unique solution to (\ref{eq:1rsb_av}), while the 1RSB phase is defined by the appearance of a non-trivial solution, which is the relevant one to describe the typical properties of (\ref{eq:prob_auxiliary}).
This non-trivial solution can appear continuously: at the RS/RSB transition, the trivial RS solution becomes unstable and the iterations of equation (\ref{eq:1rsb_av}) flow toward another fixed-point.
In the next section, we explain how to detect such a continuous RS/RSB phase transition by a stability analysis.
Note however that another scenario is possible, and wouldn't be detected by this stability analysis: the trivial RS solutions stays stable, but another non-trivial solution appears discontinuously. 
In that case the study of the local instability of the trivial fixed-point do not allow to detect the presence of an RS/RSB transition. This scenario is for instance observed on random Constraint Satisfaction Problems (e.g. on the $q$-coloring on random graphs). 
\\

\subsection{Stability of the RS solution:} 
We now explain how to analyze the stability of the RS solution to equation (\ref{eq:RS_eqn}) found numerically with population dynamics. 
We will follow the analysis derived in \cite{GaDaSeZd17}, appendix B.

Starting from the trivial solution (\ref{eq:trivial_fp}), one assumes that the distributions $P$ in the support of $\mathcal{P}^{\rm 1rsb}$ are close to Dirac deltas $\delta(\mu,\bar{\mu})$.
One can measure their distance w.r.t the Diracs by computing their first two moments:
\begin{align}
M(T)&=\int {\rm d}P(\mu)\left(\mu(T)-\bar{\mu}(T)\right)\\
V(T,T')&=\int {\rm d}P(\mu)\left(\mu(T)-\bar{\mu}(T)\right)\left(\mu(T')-\bar{\mu}(T')\right)
\end{align}
{\it A remark on the first moment:}\\
Note that the average of the first moment is always equal to $0$. 
Indeed, one can check that the random variable $\bar{\mu}[P]=\int{\rm d}P(\mu)\mu$ is distributed according to the RS probability $\mathcal{P}^{\rm rs}$.
Therefore, once averaged over the disorder: $P\sim \mathcal{P}^{\rm 1rsb}$, one obtains $\mathbb{E}_{P\sim \mathcal{P}^{\rm 1rsb}}[M(T)]=0$.\\
However, it does not means that the random variable $M(\tau)$ is always equal to $0$. 

One typically has $M(T)=0$ for problems with more symmetries, such as the graph $q$-coloring or hyper-graph bi-coloring, where by invariance under permutation of colors, the marginals should be equal to the flat distribution over colors. In such cases, once averaged over clusters, one has the following identity: $\bar{\mu}[P]=\int{\rm d}P(\mu)\mu=\bar{\mu}=1/q$. 
But in the case of epidemics there is no such symmetry and a priori $M(T)$ has a non-trivial distribution. 
\\

\subsection{Recursion on the two first moments:}
The goal is to find a recursion relation for the first and second moment.
One can do it by expanding linearly the BP equation around the RS fixed-point, considering that the BP messages are written $\mu_{i\to j}=\bar{\mu}_{i\to j}+\epsilon_{i\to j}$. Expanding in the $\epsilon_{i\to j}$'s leads to:
\begin{align}
\epsilon_{i\to j}(T_{ij})=\sum_{m\in\partial i\setminus j}\sum_{T_{mi}}\epsilon_{m\to i}(T_{mi})\left.\frac{\partial f(T_{ij})}{\partial\mu_{mi}(T_{mi})}\right|_*
\end{align}
where the derivative is evaluated at the RS fixed-point:
$$
\left.\frac{\partial f(T_{ij})}{\partial\mu_{mi}(T_{mi})}\right|_*=\left.\frac{\partial f(T_{ij})}{\partial\mu_{mi}(T_{mi})}\right|_{\mu_{k\to i}=\bar{\mu}_{k\to i}} \  .
$$
One injects this expansion in $M(T)=\int{\rm d}P(\mu)\epsilon(T)$ and in $V(T,T')=\int{\rm d}P(\mu)\epsilon(T)\epsilon(T')$, with $P(\mu)$ obeying the recursion relation (\ref{eq:1rsb_ongraph}).
By noticing that the normalization $Z_{i\to j}$ cancels at the lowest order with the re-weighting term $z_{i\to j}^x$, one obtains the following relations:
\begin{align}
\label{eq:1st_mom}
M^{(t+1)}(T)&=\sum_{i=1}^{d-1}\sum_{T_i}\left.\frac{\partial f(T)}{\partial\mu_i(T_i)}\right|_*M_i^{(t)}(T_i)\\
\label{eq:2nd_mom}
V^{(t+1)}(T,T')&=\sum_{i=1}^{d-1}\sum_{T_i,T_i'}\left.\frac{\partial f(T)}{\partial\mu_i(T_i)}\right|_*\left.\frac{\partial f(T')}{\partial\mu_i(T_i')}\right|_*V_i^{(t)}(T_i,T_i')
\end{align}
where for each $i\in\{1,\dots,d-1\}$, $M_i,V_i$ are the first two moments of the distribution $P_i$, with the $P_1,\dots,P_d$ sampled i.i.d. from $\mathcal{P}^{\rm 1rsb}$.
We have added the superscript $(t)$ to emphasize the iterative resolution of these self-consistency equations with population dynamics.

Therefore, to study the stability of the RS solution, one needs to attach to the RS population of BP messages $\{\mu_i\}_{i\in\{1,\dots,\mathcal{N}\}}$, a population of variances: $\{(\mu_1,M_1,V_1),\dots,(\mu_{\mathcal{N}},M_{\mathcal{N}},V_{\mathcal{N}})\}$. A new element $(\mu_i,M_i,V_i)$ is computed from (\ref{eq:BP_equations}) and (\ref{eq:1st_mom},\ref{eq:2nd_mom}).
\\

To assess the stability of the fixed-point, the rate of growth of the variance can be computed as:
\begin{align}
\label{eq:Delta_M}
\Delta_{M}(t)&=\frac{1}{\mathcal{N}}\sum_{i=1}^\mathcal{N}\sqrt{\frac{1}{q}\sum_{T}(M_i^{(t)}(T))^2} \\
\Delta_{V}(t)&=\frac{1}{\mathcal{N}}\sum_{i=1}^\mathcal{N}\frac{1}{q^2}\sum_{T,T'}V_i^{(t)}(T,T')
\end{align}
with $q=\mathcal{T}^4$ the size of the variable $T$ gathering the four times (remember that $T_{ij}=(\tau_i^{(j)},\tau_j^{(i)},t_i^{(j)},t_j^{(i)})$).

As already noted in the previous sub-section, studying the first moment $M(T)$ is sufficient for the distribution (\ref{eq:prob_auxiliary}), since it does not exhibit any particular symmetry forcing $M(T)$ to be equal to $0$. 
We therefore concentrate our results on the stability parameter $\Delta_t^{(M)}$ (it is the parameter shown in Figure~\ref{fig:RS_instab_twoIC}).

\subsection{Stability criterion and initial condition}
\label{appsubsec:stab_IC}
For an initial condition $\Delta_M(t=0)$ sufficiently small, we observe an exponential decay of the stability parameter $\Delta_M(t)$:
$$
\Delta_M(t)\sim e^{\delta t}
$$
The exponential rate $\delta$ can be extracted from a linear fit of $\ln(\Delta_M(t))$, and gives a criterion for characterizing the instability of the RS fixed-point:
\begin{itemize}
	\item For $\delta<1$, the RS fixed-point is stable under a perturbation towards the space of 1RSB solutions
	\item For $\delta>1$, the RS fixed-point is unstable towards the space of 1RSB solutions. This instability unveils the presence of a non-trivial solution to the 1RSB equations (\ref{eq:1rsb_av}), characteristic of an RSB phase.
\end{itemize}
Once a RS fixed-point is found, i.e. after a given number of iterative steps for the numerical resolution of (\ref{eq:RS_eqn}, we switch-on the perturbation.
For the initialization, we choose for each element $i\in\{1,\dots \mathcal{N}\}$ of the population to following perturbation, for all $T$:
\begin{align}
	M_i(T) = \sigma_0 \mu_i(T) 
\end{align}
with $\sigma_0\ll 1$ a parameter, in such a way that the perturbation $M_i(T)$ is zero on each component for which the associated BP message $\mu_i(T)$ is zero.
The variance of the perturbation at initial times is therefore $\sigma_0$ times smaller than the variance of the BP messages at the fixed-point:
\begin{align}
\Delta_M(0)=\sigma_0\Delta_\mu^*
\end{align}

\subsection{Recursion for the first moment: simplifications}
One can use exactly the same tricks to decrease the size of the moments $M(T)$ as the ones used to decrease the size of the BP messages $\mu(T)$.\\
Re-writing the BP equation (\ref{eq:BP_equations}) as: 
\begin{align}
\begin{aligned}
\mu_{i\to \Psi_j}(T_{ij})&=\mu_{ij}(T_{ij})=\frac{\omega_{ij}(T_{ij})}{z_{ij}}  \\
\omega_{ij}(T_{ij})&= \sum_{\{T_{ki}\}_{k\in\partial i\setminus j}}\Psi(\{T_{il}\}_{l\in\partial i})\prod_{k\in\partial i\setminus j}\mu_{ki}(T_{ki})   \\
z_{ij}&=\sum_{T_{ij}}\omega_{ij}(T_{ij})
\end{aligned}
\end{align}
One can re-write the recursion relation (\ref{eq:1st_mom}) as:
\begin{align}
\begin{aligned}
M_{ij}(T_{ij})&=\frac{1}{z_{ij}}\sum_{m\in\partial i\setminus j}\omega_{ij}^{(m)}(T_{ij}) \\
&- \frac{\omega_{ij}(T_{ij})}{z_{ij}^2}\sum_{m\in\partial i \setminus j}z_{ij}^{(m)}
\end{aligned}
\end{align}
With:
\begin{align}
\begin{aligned}
\omega_{ij}^{(m)}(T_{ij})&=\sum_{\{T_{ki}\}_{k\in\partial i\setminus j}}\Psi(\{T_{il}\}_{l\in\partial i})\\
&\times M_{mi}(T_{mi})\times \prod_{k\in\partial i\setminus \{j,m\}}\mu_{ki}(T_{ki})\\
z_{ij}^{(m)} &= \sum_{T_{ij}}w_{ij}^{(m)}(T_{ij})
\end{aligned}
\end{align}
Just as we did for the BP messages, one can define the auxiliary messages:
\begin{align}
\begin{aligned}
\tilde{\omega}^{(m)}_{ij}(\tau_i,\sigma_{ji},t_i,t_j)&=\omega^{(m)}_{ij}(\tau_i,\tau_j,t_i,t_j)
\end{aligned}
\end{align}
$\forall$ $\tau_j$ such that $\sigma_{ji}=1+{\rm sgn}[\tau_j-\tau_i+s_{ji}]$, and:
\begin{align}
\begin{aligned}
\tilde{M}_{ij}(\sigma_{ij},\tau_j,c_{ij},t_j)&=\sum_{t_i}a(t_j-t_i-c_{ij})\\
\times \sum_{\tau_i}M_{ij}(\tau_i,\tau_j,t_i,t_j)&\mathbb{I}[\sigma_{ij}=1+{\rm sgn}[\tau_i-\tau_j+s_{ij}]]
\end{aligned}
\end{align}
And obtain a recursion relation for these messages. 
These equations are identical to the BP equations (\ref{eq:BP_equations_simplif}), except that we replace $\tilde{\mu}_{mi}$ by $\tilde{M}_{mi}$ in the equation for $\tilde{\omega}_{ij}^{(m)}$. More precisely, we obtain the following equations:
\begin{widetext}
	\begin{align}
	\begin{aligned}
	\label{eq:stab_equations_simplif}
	\tilde{M}_{i\to \Psi_j}(\sigma_{ij},\tau_j^{(i)},c_{ij},t_j^{(i)}) &= \sum_{t_i^{(j)}}a(t_j^{(i)}-t_j^{(i)}-c_{ij})\sum_{\tau_i^{(j)}}N_{\Psi_i\to j}(\tau_i^{(j)},\sigma_{ji}=1+\text{sgn}(\tau_j^{(i)}-\tau_i^{j}+s_{ji}),t_i^{(j)},t_j^{(i)})\\
	&\times\mathbb{I}[\sigma_{ij}=1+\text{sgn}(\tau_i^{(j)}-\tau_j^{(i)}+s_{ij})] 
	\end{aligned}
	\end{align}
	where
	\begin{align}
	N_{\Psi_i\to j}(\tau_i^{(j)},\sigma_{ji},t_i^{(j)},t_j^{(i)}) &=\frac{1}{z_{ij}}\sum_{m\in\partial i\setminus j}\tilde{\omega}_{ij}^{(m)}(\tau_i^{(j)},\sigma_{ji},t_i^{(j)},t_j^{(i)}) - \frac{\tilde{\omega}_{ij}(\tau_i^{(j)},\sigma_{ji},t_i^{(j)},t_j^{(i)})}{z_{ij}^2}\sum_{m\in\partial i \setminus j}z_{ij}^{(m)}
	\end{align}
	and:
	\begin{align}
	\begin{aligned}	
	\tilde{\omega}^{(m)}_{\Psi_i\to j}(\tau_i^{(j)},\sigma_{ji},t_i^{(j)},t_j^{(i)})&=\gamma(t_i^{(j)})\xi(\tau_i^{(j)},t_i^{(j)})(A^{(m)}_1(\tau_i^{(j)},\sigma_{ji},t_i^{(j)},t_j^{(i)})-\phi(t_i^{(j)})A^{(m)}_0(\tau_i^{(j)},\sigma_{ji},t_i^{(j)},t_j^{(i)}))  \\
	\end{aligned}
	\end{align}
	with: 
	\begin{align}
	\begin{aligned}
	&A^{(m)}_c(\tau_i^{(j)},\sigma_{ji},t_i^{(j)},t_j^{(i)})\\
	&=a(t_i^{(j)}-t_j^{(i)}-c)\delta_{x_i^0,I}\delta_{\tau_i^{(j)},0}\left(\sum_{\sigma_{ki}=0}^2\tilde{M}_{m\to\Psi_i}(\sigma_{ki},\tau_i^{(k)},c_{ki}=c,t_i^{(k)})\right)\prod_{k\in\partial i\setminus \{j,m\}}\left(\sum_{\sigma_{ki}=0}^2\tilde{\mu}_{k\to\Psi_i}(\sigma_{ki},\tau_i^{(k)},c_{ki}=c,t_i^{(k)})\right) \\
	&+a(t_i^{(j)}-t_j^{(i)}-c)\delta_{x_i^0,S}\delta_{\sigma_{ij}\in\{1,2\}}\left(\sum_{\sigma_{ki}=1}^2\tilde{M}_{m\to\Psi_i}(\sigma_{ki},\tau_i^{(k)},c_{ki}=c,t_i^{(k)})\right)\prod_{k\in\partial i\setminus \{j,m\}}\left(\sum_{\sigma_{ki}=1}^2\tilde{\mu}_{k\to\Psi_i}(\sigma_{ki},\tau_i^{(k)},c_{ki}=c,t_i^{(k)})\right) \\
	&-a(t_i^{(j)}-t_j^{(i)}-c)\delta_{x_i^0,S}\delta_{\tau_i^{(j)}<\mathcal{T}}\delta_{\sigma_{ij}=2}\left(\tilde{M}_{m\to\Psi_i}(\sigma_{ki}=2,\tau_i^{(k)},c_{ki}=c,t_i^{(k)})\right)\prod_{k\in\partial i\setminus \{j,m\}}\left(\tilde{\mu}_{k\to\Psi_i}(\sigma_{ki}=2,\tau_i^{(k)},c_{ki}=c,t_i^{(k)})\right) \\
	\end{aligned}
	\end{align}	
\end{widetext}

\section{Instability study on finite-size instances}
\label{app:BP_stab}
In this appendix, we present the stability analysis of the BP equations on a single instance on a given contact graph $G=(V,E)$, and for a given realization of the spreading process.

\subsection{BP equations for the posterior probability on a single instance}
We start from the posterior probability distribution (\ref{eq:posterior}), that we re-write here in terms of the factor functions $\psi(t_i,\underline{t}_{\partial i})$:
\begin{align}
\begin{aligned}
	P(\underline{t}|\mathcal{O}) &= \frac{P(\underline{t}P(\mathcal{O}|\underline{t}))}{P(\mathcal{O})} \\
	&=\frac{1}{P(\mathcal{O})}\prod_{i\in V}\psi(t_i\underline{t}_{\partial i})\rho(o_i,t_i)
\end{aligned}
\end{align}
with $\rho(o_i,t_i)$ given in (\ref{eq:observations}) and with:
$$
\psi(t_i\underline{t}_{\partial i}) = \sum_{\mathcal{D}_i}\psi^*(t_i\underline{t}_{\partial i};\mathcal{D}_i)
$$
where $\psi^*$ is given in (\ref{eq:constraint_infection_times}).
The probability $P(\mathcal{O})$ plays the role of a normalization. 
As in the case of the joint probability distribution (\ref{eq:joint_short_loops}), the {\it posterior} probability distribution contains short loops that prevent from a direct application of a message-passing approach.

In order to cure these short loops, and following the steps of \cite{AlBrDaLaZe14}, we can introduce auxiliary inferred infection times $t_i^{(j)}, t_j^{(i)}$ on each edges of the contact graph $(i,j)\in E$, similarly as we did in appendix \ref{appsubsec:aux_vars_joint}, with the constraint:
\begin{align}
t_i^{(j)}=t_i \ , \  \forall j\in\partial i 
\end{align}
The probability distribution on the auxiliary variables becomes:
\begin{align}
 P(\{t_i^{(j)},t_j^{(i)}\}_{i,j}\in E) = \frac{1}{P(\mathcal{O})}\prod_{i\in V}\chi(\{t_i^{(l)}t_l^{(i)}\}_{l\in\partial i})
\end{align}
where (for any $j\in\partial i$):
\begin{align}
\chi(\{t_i^{(l)}t_l^{(i)}\}_{l\in\partial i}) = \psi(t_i^{(j)},\underline{t}_{\partial i}(i))\prod_{l\in\partial i\setminus j}\delta_{t_i^{(l)},t_i^{(j)}}
\end{align}
The contact graph associated with the distribution on the auxiliary variables is identical to the one obtained in section \ref{appsubsec:aux_vars_joint}, cf Figure \ref{fig:factor_graph}, right panel. The variable nodes representing auxiliary variable live on the edges of the original contact graph $G=(V,E)$, while the factor nodes representing the factor functions $\chi$ live on the vertices of the contact graph. The systematic short loops have been removed, and 
one can define the variable-to-factor BP messages $\mu_{i\to\chi_j}$, satisfying the following BP equations:
\begin{align}
\label{eq:BP_onegraph_app}
\begin{aligned}
\mu_{i\to\chi_j}(t_i^{(j)},t_j^{(i)}) &= \frac{1}{z_{i\to\chi_j}}\sum_{\{t_i^{(l)},t_l^{(i)}\}_{l\in\partial i \setminus j}}\chi(\{t_i^{(l)},t_l^{(i)}\}_{l\in\partial i}) \\
&\times\prod_{l\in\partial i\setminus j}\mu_{l\to\chi_i}(t_l^{(i)},t_i^{(l)})
\end{aligned}
\end{align}
The size of the BP message $\mu_{i\to \chi_j}$ can be decreased by introducing auxiliary messages, similarly as it was done in section \ref{appsubsec:simplifs_BP_joint}.

\subsection{Stability analysis}
On a given instance $G,\mathcal{O}$, a solution to the BP equation (\ref{eq:BP_onegraph_app}) can be found numerically with an iterative procedure. 
Let $\{\mu_{i\to\chi_j}^*,\mu_{j\to\chi_i}^*\}_{(i,j)\in E}$ be a fixed-point of the BP equations.
Its stability can be assessed by introducing a small perturbation $\epsilon_{i\to \chi_j}$ to each messages, and study its evolution under the updates of the iterative procedure.
A linear expansion of the BP equation gives the update rule for the perturbation terms:
\begin{align}
\label{eq:stab_onegraph_update}
\begin{aligned}
\epsilon_{i\to \chi_j}(t_i^{(j)},t_j^{(i)}) &= \sum_{m\in\partial i\setminus j}\sum_{t_m^{(i)},t_i^{(m)}}\epsilon_{m\to \chi_i}(t_m^{(i)},t_i^{(m)})\\
\times&\left.\frac{\partial f^{\rm BP}(t_i^{(j)},t_j^{(i)})}{\partial \mu_{m\to\chi_i}(t_m^{(i)},t_i^{(m)})}\right|_*
\end{aligned}
\end{align}
where the function $f^{\rm BP}$ is referring to the BP equation (\ref{eq:BP_onegraph_app}):
$$
\mu_{i\to\chi_j} = f^{\rm BP}(\{\mu_{l\to\chi_i}\}_{l\in\partial i\setminus j})
$$
and where the derivative of $f^{\rm BP}$ is evaluated at the fixed-point:
$$
\left.\frac{\partial f^{\rm BP}(t_i^{(j)},t_j^{(i)})}{\partial \mu_{m\to\chi_i}(t_m^{(i)},t_i^{(m)})}\right|_* = \left.\frac{\partial f^{\rm BP}(t_i^{(j)},t_j^{(i)})}{\partial \mu_{m\to\chi_i}(t_m^{(i)},t_i^{(m)})}\right|_{\mu_{i\to\chi_j}=\mu_{i\to\chi_j}^*}
$$
Note the similarity of the update equation (\ref{eq:stab_onegraph_update}) with the update rule for the perturbation of the RS fixed-point (\ref{eq:1st_mom}), with the difference in the BP function $f^{\rm BP}$ (eq. (\ref{eq:BP_onegraph_app})) and $f$ (eq. (\ref{eq:BP_equations})).

The variance of the perturbations over all messages (cf equation (\ref{eq:stab_onegraph})):
$$
\Delta(t;G,\mathcal{O})=\frac{1}{2|E|}\sum_{i\in V}\sum_{j\in\partial i}\sqrt{\frac{1}{q}\sum_{T\in\chi}(\epsilon_{i\to j}(T))^2}
$$
is plotted in section \ref{subsec:instability_BP}, Figure \ref{fig:BP_instab_fit} (inset, right panel).

As for the stability analysis of the RS fixed-point, the variance $\Delta(t,G,\mathcal{O})$ displays an exponential decay:
$$
\Delta(t,G,\mathcal{O})\sim e^{\delta(G,\mathcal{O})t}
$$
from which the exponential rate $\delta(G,\mathcal{O})$ can be extracted by a linear fit.

Despite the similarities between the stability study of the RS fixed-point (appendix \ref{app:RS_stab}) and the stability study of BP updates on a graph, one shall emphasize that the two analysis are aiming at different purposes:
\begin{itemize}
	\item The RS stability analysis is made on average over instances $G,\mathcal{O}$, {\it in the large size limit}, and unveils a phase transition towards an RSB phase
	\item The BP analysis is made for one instance, the stability parameter $\delta(G,\mathcal{O})$ being averaged over a finite sample of finite-size instances. It detects the instability of the BP fixed-point found on {\it finite-size instances}, and cannot be used by itself to conclude on the presence of an RSB phase.
\end{itemize}
\end{document}